\documentclass[%
reprint,
 amsmath,amssymb,
pra,
]{revtex4-2}

\usepackage{graphicx}
\usepackage{dcolumn}
\usepackage{bm}
\usepackage{amsmath}
\usepackage{nccmath}
\usepackage[dvipdfmx]{hyperref}
\usepackage[dvipdfmx]{xcolor}
\usepackage{mathtools}
\usepackage{natbib}

\begin{document}

\preprint{APS/123-QED}

\title{The origin of correlation between mass and angle in quasi-fission}

\author{S. Amano, Y. Aritomo, S. Ishizaki, M. Okubayashi and S. Okugawa\\
\scriptsize{Kindai University Higashi-Osaka, Osaka 577-8502, Japan}\\
\scriptsize{e-mail: 2144340401y@kindai.ac.jp}\\
}

\date{\today}

\begin{abstract}
Mass-angle distribution (MAD) measurement of heavy and superheavy element fragmentation reactions is one of the powerful tools for investigating the mechanism of fission and fusion process. MAD shows a strong correlation between mass and angle when the quasi-fission event is dominant. It has characteristic that appears diagonal correlation as long as the quasi-fission event is dominant. This diagonal correlation could not be reproduced in previous our model before the introduction of the parameters.

In this study, we systematically evaluate the unknown model parameters contained in our model and clarify those model parameters to reproduce the diagonal correlation that appears in MAD. Using a dynamical model based on the fluctuation dissipation theorem that employs Langevin equations, we calculate MADs of two reaction systems $^{48}$Ti+$^{186}$W and $^{34}$S+$^{232}$Th which are dominated by quasi-fission. We were able to clarify the effects of unknown model parameters on the MAD. In addition, we identified the values of model parameters that can reproduce the correlation between mass and angle. As a result, it was found that the balance of tangential friction and moment of inertia values is important for the correlation between mass and angle.
\end{abstract}

\maketitle


\section{Introduction}
Currently, superheavy element synthesis is limited to the success of element Z = 118. The production of new superheavy elements will foot in the 8th period of the periodic table, and will provide insight into the existence of  island of stability. In addition, neutron-rich nuclei far from the valley of stability are important for understanding the r-process in astronomical nuclear physics. The supernova explosion is one of the origins of the elements along with the evolution of the stars.
During a supernova explosion, a large amount of new elements are created and spread around. 
After those elements create the new star, the new star eventually die, leading to an explosion again.
The elements in the universe and the solar system is created through such an evolution cycle of stars and elements. 
Superheavy elements and neutron-rich nuclei create in the process of that element synthesis scenario.
Production of those nuclei are significant to elucidate element synthesis scenario but, extremely difficult.
In recent years, the method using the nucleon transfer reaction for the synthesis of new superheavy elements and neutron-rich nuclei has been proposed \cite{karp17,zagr082}, but a quantitative prediction model has not been established. The mechanism of the nucleon transfer reaction has been studied for many years in many models, including semiclassical models such as  GRAZING \cite{wint994} and Langevin \cite{karp17}, and microscopic models such as time-depent Hartree-Fock (TDHF) \cite{ayik017}.

In this study, we analyze the relationship between mass and angle obtained by the nucleon transfer reaction.
Mass-angle distribution (MAD) is the emission angle on the vertical axis and the mass ratio $M_{R}$ on the horizontal axis.
$M_{R}$ is represented by 
$M_{R}=\frac{A_ {i = 1,2}}{A_{1} + A_{2}}$. 
$A_{1}$ and $A_{2}$ represent the mass number of the projectile-like fragment and the target-like fragment, respectively. 
When the projectile nucleus collides with the target nucleus, the projectile nucleus receives nucleons from the target nucleus while rubbing around the target nucleus, increases the mass number, and finally leaves from the target nucleus in a certain direction. 
There is a correlation between the number of nucleon transferred (mass) and the emission angle, and the characteristic is depends on the projectile nucleus and the target nucleus. 
By analyzing this correlation, we are trying to elucidate the fusion process. MAD shows a strong correlation between mass and angle when the quasi-fission reaction is dominant \cite{riez13}. 
This correlation could not be reproduced with the dynamical model developed in the previous research. In this study, we evaluated MADs of the two reaction systems $^{48}$Ti + $^{186}$W and $^{34}$S + $^{232}$Th dominated quasi-fission using a dynamical model.
This paper contain as follows Sec. 2 describes the details of the model framework. Sec. 3 discusses the parameter dependence of MAD. The last section describes conclusion.
\section{Framework}
\subsection{Potential energy surface}
We adopt the dynamical model which similar to unified model \cite{zagr071}. First, the initial stage of the nucleon transfer reactions consists of two parts: (1) the system is placed in the ground state of the projectile and target because the reaction proceeds is too fast for nucleons to reconfigure a single particle state 
(2) The part where the system relaxes to the ground state of the entire composite system which changes the potential energy surface to an adiabatic one. 
Therefore, we consider the time evolution of potential energy from the diabatic one $V_{diab}\left(q\right)$ to adiabatic one $V_{adiab}\left(q\right)$. 
Here, $q$ denotes a set of collective coordinates representing nuclear deformation. The diabatic potential is calculated by a folding procedure using effective nucleon-nucleon interaction \cite{zagr05,zagr071,zagr072}. 
However, the adiabatic potential energy of the system is calculated using an extended two-center shell model \cite{zagr072}. 
Then, we connect the diabatic and the adiabatic potentials with a time-dependent weighting function as follows:
\begin{eqnarray}
&&V=V_{diab}\left(q\right)f\left(t\right)+V_{adiab}\left(q\right)\left[1-f\left(t\right)\right], \nonumber \\
&&f\left(t\right)=\exp{\left(-\frac{t}{\tau}\right)}.
\label{pot}
\end{eqnarray}
Where $t$ is the interaction time and $f\left(t\right)$ is the weighting function included the relaxation time $\tau$. 
We use the relaxation time $\tau=10^{-21}s$ proposed in \cite{bert78,cass83,diaz04}. 
We use the two-center parameterization \cite{maru72,sato78} as coordinates to represent nuclear deformation.
To solve the dynamical equation numerically and avoid the huge computation time, 
we strictly limited the number of degrees of freedom and employ three parameters as follows: 
$z_{0}$ (distance between the centers of two potentials), 
$\delta$ (deformation of fragment), and $\alpha$ (mass asymmetry of colliding nuclei);
$\alpha=\frac{\left(A_{1}-A_{2}\right)}{\left(A_{1}+A_{2}\right)}$, 
where $A_{1}$ and $A_{2}$ not only stand for the mass numbers of the target and projectile, 
respectively \cite{zagr05,arit04} but also are then used to indicate mass numbers of the two fission fragments. 
As shown in Fig.~1 in Ref. \cite{maru72}, the parameter $\delta$ is defined as $\delta=\frac{3\left(a-b\right)}{\left(2a+b\right)}$, where $a$ and $b$ represent the half  length of the ellipse axes in the $z_{0}$ and $\rho$ directions, respectively. 
We assume that each fragment has the same deformation as a first step. 
In addition, we use scaling to save computation time and use the coordinate $z$ defined as $z=\frac{z_{0}}{\left(R_{CN}B\right)}$, where $R_{CN}$ denotes the radius of the spherical compound nucleus and the parameter $B$ is defined as $B=\frac{\left(3+\delta\right)}{\left(3-2\delta\right)}$.

\subsection{Dynamical equations}
\begin{figure*}[t]
\begin{center}
\includegraphics[scale=0.45]{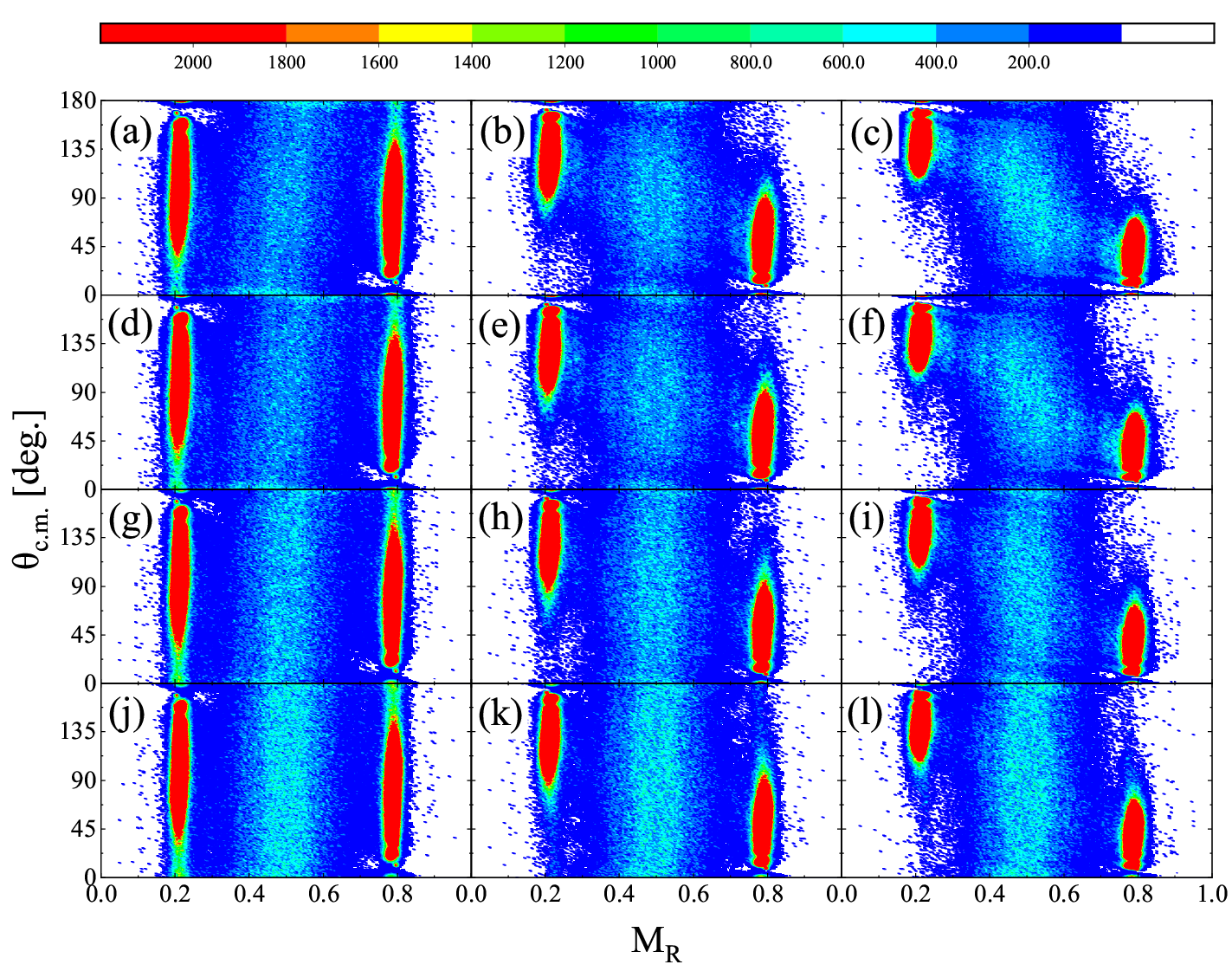}
\end{center}
\caption{The calculation MADs for $^{48}$Ti+$^{186}$W at E$_{\rm c.m.}$=187.87MeV and their dependence on $f_{ina}$ and $\gamma_t^0$ parameters in the range of $f_{ina}$ =1.5-5.0 and $\gamma_t^0$=0.5-10. 
The result of (a)(d)(g)(j), (b)(e)(h)(k) and (c)(f)(i)(l) adopt $f_{ina}$ =1.5, and $f_{ina}$ =3.0 and $f_{ina}$ =5.0, respectively.
(a)-(c), (d)-(f), (g)-(i) and (j)-(l) adopt $\gamma_t^0$=0.5,  $\gamma_t^0$=1.0,  $\gamma_t^0$=5.0 and $\gamma_t^0$=10, respectively. 
}\label{fig1}
\end{figure*}
\begin{figure*}[t]
\begin{center}
\includegraphics[scale=0.45]{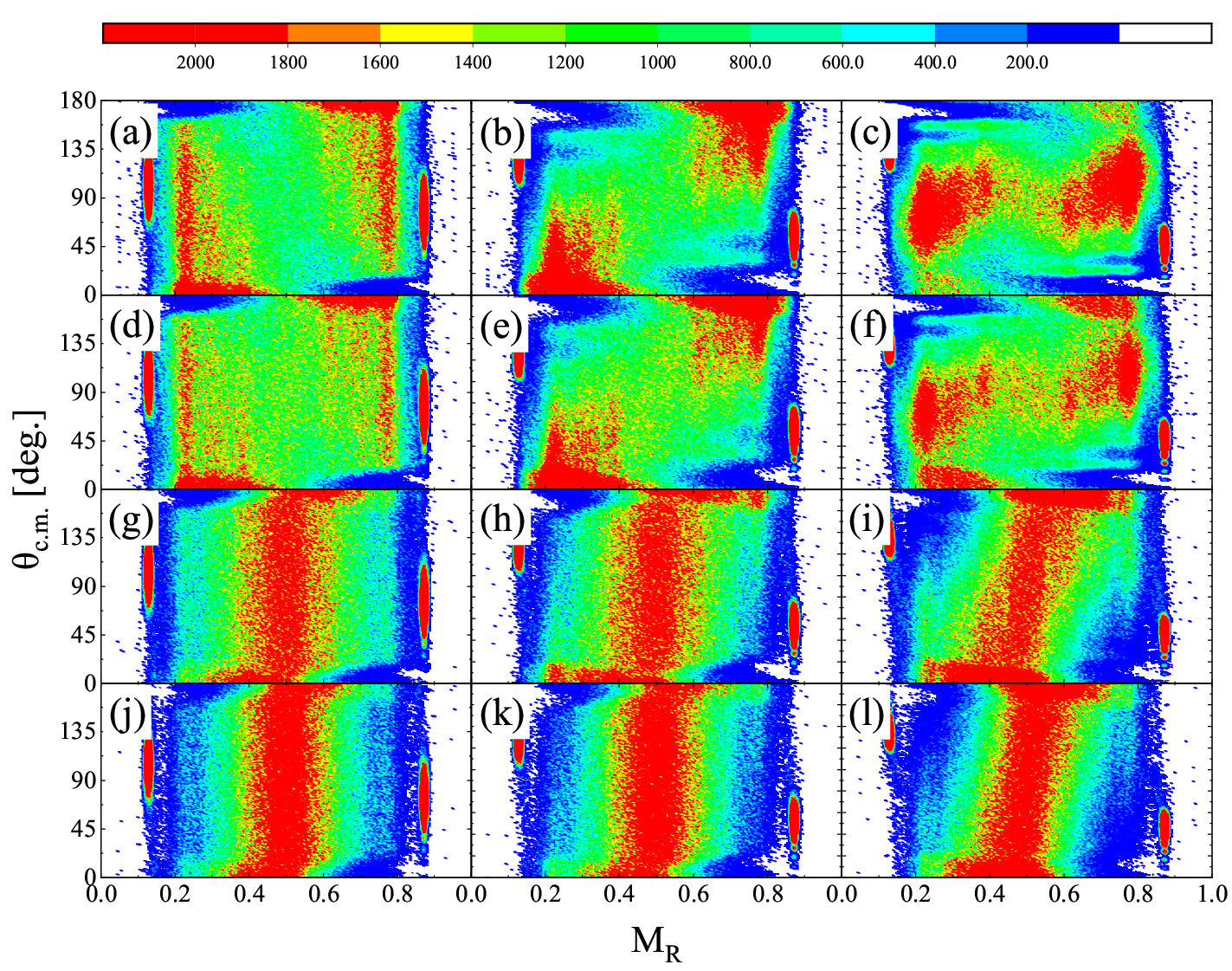}
\end{center}
\caption{The calculation MADs for $^{34}$S+$^{232}$Th at E$_{\rm c.m.}$=168.75MeV and their dependence on $f_{ina}$ and $\gamma_t^0$ parameters in the range of $f_{ina}$ =1.5-5.0 and $\gamma_t^0$=0.5-10.
The result of (a)(d)(g)(j), (b)(e)(h)(k) and (c)(f)(i)(l) adopt $f_{ina}$ =1.5, and $f_{ina}$ =3.0 and $f_{ina}$ =5.0, respectively.
(a)-(c), (d)-(f), (g)-(i) and (j)-(l) adopt $\gamma_t^0$=0.5,  $\gamma_t^0$=1.0,  $\gamma_t^0$=5.0 and $\gamma_t^0$=10, respectively. 
}\label{fig2}
\end{figure*}
We perform trajectory calculations of the time-dependent unified potential energy \cite{zagr05,zagr071,arit04} by Langevin equation. 
We start trajectory calculations from a sufficiently long distance between both nuclei \cite{arit04}. 
So, we use the model which takes into account the nucleon transfer for slightly separated nuclei \cite{zagr05}. 
Process for the separated nucleon transfer use the procedure described in Refs. \cite{zagr05,zagr071}.
When both nuclei has been changed the mononucleus state that window of the contact nuclei is sufficiently opened, the evolution process of the mass asymmetry parameter $\alpha$ switches from the master equation to Langevin equation according to the procedure described in Ref. \cite{arit04}.
We use the multidimensional Langevin equation \cite{zagr05,arit04,arit09} unified following:
\begin{gather}
\frac{dq_{i}}{dt}=\left(m^{-1}\right)_{ij}p_{j}, \nonumber \\
\frac{dp_{i}}{dt}=-\frac{\partial V}{\partial q_{i}}-\frac{1}{2}\frac{\partial}{\partial q_{i}}\left(m^{-1}\right)_{jk}p_{j}p_{k}-\gamma_{ij}\left(m^{-1}\right)_{jk}p_{k} \nonumber \\
+g_{ij}R_{j}\left(t\right), \nonumber \\
\frac{d\theta}{dt}=\frac{\ell}{\mu_{R}R^{2}}, \qquad 
\frac{d\varphi_{1}}{dt}=\frac{L_{1}}{\Im_{1}}, \qquad
\frac{d\varphi_{2}}{dt}=\frac{L_{2}}{\Im_{2}}, \nonumber \\
%
\frac{d\ell}{dt}=-\frac{\partial V}{\partial\theta}-\gamma_{tan}\left(\frac{\ell}{\mu_{R}R^{2}}-\frac{L_{1}}{\Im_{1}}a_{1}-\frac{L_{2}}{\Im_{2}}a_{2}\right)R
\nonumber \\
+Rg_{tan}R_{tan}\left(t\right), \nonumber \\
\frac{dL_{1}}{dt}=-\frac{\partial V}{\partial\varphi_{1}}+\gamma_{tan}\left(\frac{\ell}{\mu_{R}R^{2}}-\frac{L_{1}}{\Im_{1}}a_{1}-\frac{L_{2}}{\Im_{2}}a_{2}\right)a_{1}  \nonumber \\
-a_{1}g_{tan}R_{tan}\left(t\right), \nonumber \\
\frac{dL_{2}}{dt}=-\frac{\partial V}{\partial\varphi_{2}}+\gamma_{tan}\left(\frac{\ell}{\mu_{R}R^{2}}-\frac{L_{1}}{\Im_{1}}a_{1}-\frac{L_{2}}{\Im_{2}}a_{2}\right)a_{2}
\nonumber \\
-a_{2}g_{tan}R_{tan}\left(t\right).
\end{gather}
The collective coordinates $q_{i}$ represent $z, \delta$, and $\alpha,$ the symbol $p_{i}$ denotes  momentum conjugated to $q_{i}$, and $V$ is the multidimensional potential energy. 
: The symbols $\theta$ and $\ell$ indicates the relative  orientation of nuclei and relative angular momentum respectively. $\varphi_{1}$ and $\varphi_{2}$ stand for the rotation angles of the nuclei in the reaction plane (their moment of inertia and angular momenta are $\Im_{1,2}$ and $L_{1,2}$, respectively), $a_{1,2}=\frac{R}{2}\pm\frac{\left(R_{1}-R_{2}\right)}{2}$ is the distance from the center of the fragment to the middle point between the nuclear surfaces, and $R_{1,2}$ is the nuclear radii. The symbol $R$ is distance between the nuclear centers. 
The total angular momentum $L=\ell+L_{1}+L_{2}$ is preserved. The symbol $\mu_{R}$ is reduced mass, and $\gamma_{tan}$ is the tangential friction force of the colliding nuclei. 
Here, it is called sliding friction. 
The phenomenological nuclear friction forces for separated nuclei are expressed in terms of $\gamma_{tan}^{F}$ for sliding friction using the Woods-Saxon radial form factor described in Refs. \cite{zagr05,zagr071}. 
The sliding friction are described as $\gamma_{tan}=\gamma_{t}^{0}F\left(\zeta\right)$, where the radial form factor $\ F\left(\zeta\right)=\left(1+\exp^{\zeta}\right)^{-1}, \zeta=\frac{\left(\xi-\rho_{F}\right)}{a_{F}}$. $\gamma_{t}^{0}$ denote the strength of the tangential friction, respectively. $\rho_{F} \sim$ 2 fm and $a_{F} \sim$ 0.6 fm are the model parameters, and $\xi$ is the distance between the nuclear surfaces $\xi=R-R_{contact}$, where $R_{contact}=R_{1}+R_{2}$ \cite{zagr05}. 
The symbols separated by $m_{ij}$ and $\gamma_{ij}$ stand for the shape-dependent collective inertia and friction tensors elements, respectively. 
We adoped the hydrodynamic inertia tensor $m_{ij}$ in Werner-Wheeler approximation for the velocity field \cite{davi76}. The normalized random force $R_{i}\left(t\right)$ is assumed to be white noise: $\langle R_{i} (t) \rangle$=0 and $\langle R_{i} (t_{1})R_{j} (t_{2})\rangle = 2 \delta_{ij}\delta (t_{1}-t{2})$. 
According to Einstein relation, the strength of the random force $g_{ij}$ is given $\gamma_{ij}T=\sum_{k}{g_{ij}g_{jk}}$, where $T$ is the temperature of the compound nucleus calculated from the intrinstic energy of the composite system.
The adiabatic potential energy is defined as
\begin{eqnarray}
&V_\mathrm{{adiab}}\left(q,L,T\right)=V_{LD}\left(q\right)+
\frac{\hbar^{2}L\left(L+1\right)}{2\mathcal{I}\left(q\right)} \nonumber 
+V_{SH}\left(q,T\right), \nonumber \\
&V_{LD}\left(q\right)=E_{S}\left(q\right)+E_{C}\left(q\right), \nonumber \\
&V_{SH}\left(q,T\right)=E_{shell}^{0}\left(q\right)\Phi\left(T\right), \nonumber \\
&\Phi\left(T\right)=\exp\left(-\frac{E^{\ast}}{E_{d}}\right).
\end{eqnarray}
Here, $\mathcal{I}\left(q\right)$ represents the moment of inertia of the rigid body with deformation $q$. 
The centrifugal energy generated from the angular momentum $L$ of the rigid body is also taken into account. 
$V_{LD}$ and $V_{SH}$ are the potential energy of the finite range liquid drop model and the shell correction energy that takes into account temperature dependence, respectively. 
The symbol $E_{shell}^{0}$ indicates the shell correction energy at $T$=0. 
The temperature dependence factor $\Phi\left(T\right)$ is explained in Ref. \cite{arit04}, where $E^{\ast}$ indicates the excitation energy of the compound nucleus. $E^{\ast}$ is given $E^{\ast}=aT^{2}$, where a is the level density parameter. 
The shell damping energy $E_{d}$ is selected as 20 MeV. This value is given by Ignatyuk et al. \cite{igna75}. 
The symbols $E_{S}$ and $E_{C}$ stand for generalized surface energy \cite{krap79} and Coulomb energy, respectively.

In this study, the calculation MADs was performed by changing two parameters ($\gamma_t ^ 0$ and $f_ {ina}$).
$\gamma_t ^ 0$ is the tangential friction correction factor, $f_ {ina}$ is shown the correction factor in the form of $\frac {\hbar^{2}L \left(L + 1\right)}{2\mathcal{I}\left(q\right)f_{ina}}$. $\mathcal{I}\left(q\right)$ represents the moment of inertia of a rigid body. 

\section{Results}
Figure 1 shows the calculation MADs for $^{48}$Ti+$^{186}$W at E$_{\rm c.m.}$=187.87MeV and their dependence on $f_{ina}$ and $\gamma_t^0$ parameters in the range of $f_{ina}$ =1.5-5.0 and $\gamma_t^0$=0.5-10. For example, in Fig. 1(k), there is no correlation in MAD, which means that the emitted nuclei are emitted in all directions of 360 degrees, which is characteristic of the dominant fusion-fission process. 
Moreover, it is considered that the situation of quasi-fission process is reproduce because Fig. 1(a) has a correlation between angle and mass.
Fig. 1(a) and 1(d) can reproduce the characteristics of the experimental values\cite{riez13} well. 
When $f_{ina}$ was large, the two bodies after contact tended to be difficult to move. It was found that when $\gamma_t ^ 0$ is large, the correlation between mass and angle is lost. 
It is considered that this is because the emission angle overs $2\pi$ and is fragments are emitted in all directions due to the dominant fusion-fission reaction. 

\par Next, Figure 2 shows the calculation results of $^{34}$S+$^{232}$Th MAD at E$_{\rm c.m.}$=168.75MeV when $\gamma_t ^ 0$ and $f_ {ina}$ are changed. 
For instance, in Fig. 2(j), there is no correlation between mass and angle to dominate fusion-fission.
On the other hand, since Fig. 2(l) has a correlation between angle and mass, it is considered that we reproduce the situation of quasi-fission process.
Fig. 2(i) and 2(l) can reproduce the characteristics of the experimental values\cite{riez13} well. 
The influence for MAD does not change even if the moment of inertia for rigid body and the tangential friction change respectively. But, the correlation originated from quasi-fission between mass and angle show when both $\gamma_t ^ 0$ and $f_ {ina}$ are large. This is unlike $^{48}$Ti+$^{186}$W case.
In this study, it was found that the tangential friction and the moment of inertia for rigid body that can reproduce the correlation between mass and angle depends on the reaction system. 
In the reaction system with a small mass asymmetry of the incident system such as $^{48}$Ti+$^{186}$W, the calculation result could reproduce the correlation originated quasi-fission like the experimental values by the small value of the tangential friction and the moment of inertia for the rigid body. 
On the other hand, the calculation result could reproduce the characteristic originate from quasi-fission when the value of the tangential friction and the moment of inertia for the rigid body large in the reaction system that the initial mass asymmetry is large such as $^{34}$S+$^{232}$Th. 
From these results, it is considered that the value of the tangential friction and the moment of inertia for reproducing the experimental value depend on the initial mass asymmetry.

\section{Conclusion}
It was found that the tangential friction and the moment of inertia for the rigid body are strongly relate to the correlation which originates from quasi-fission in the heavy element region. 
It was also found that the balance of these physical quantities shows the characteristic whether quasi-fission is dominant or fusion-fission reaction is dominant. 
It was possible to reproduce the characteristics of the experimental results which quasi-fission is dominant by changing the unknown model parameters.  
We need to calculate systematically to specify what unkown model parameters depend on. Furthermore, it is necessary to find the correlation between the values of these unknown model parameters and other physical quantities to reduce unknown parameters. 
In the future, we will investigate the dependence of tangential friction and the moment of inertia for the rigid body on mass asymmetry.

\section*{Acknowledgments}
The Langevin calculation were perfomed using the cluster computer system (Kindai-VOSTOK) which is supported by JSPS KAKENHI Grant Number 20K04003.


\begin{thebibliography}{99}
\bibitem{karp17}A.V.~Karpov and V.V.~Saiko, Phys. Rev. C {\bf 96}, 024618 (2017).
\bibitem{zagr082}V.I.~Zagrebaev, et al., Phys. Rev. C {\bf 73}, 122701 (2008).
\bibitem{wint994}A.~Winther, Phys. A {\bf 572}, 191 (1994).
\bibitem{ayik017}S.~Ayik, et al., Phys. Rev. C {\bf 96}, 024611 (2017).
\bibitem{riez13}R.du~Riez, et al., Phys. Rev. C {\bf 88}, 054618 (2013).
\bibitem{zagr05}V.~Zagrebaev and W.~Greiner, J. Phys. G {\bf 31}, 7, 825 (2005).
\bibitem{zagr071}V.~Zagrebaev and W.~Greiner, J. Phys. G {\bf 34}, 11, 2265 (2007).
\bibitem{zagr072}V.I.~Zagrebaev, et al., Phys. Part. Nuclei {\bf 38}, 469 (2007).
\bibitem{bert78}G.F.~Bertsch, Z. Phys. A {\bf 289}, 103 (1978).
\bibitem{cass83}W.~Cassing and W.~N\"{o}renberg, Nucl. Phys. A {\bf 401}, 467 (1983).
\bibitem{diaz04}A.~Diaz-Torres, Phys. Rev. C {\bf 69}, 021603 (2004).
\bibitem{maru72}J.~Maruhn and W.~Greiner, Z. Phys {\bf251}, 431 (1972).
\bibitem{sato78}K.~Sato, A.~Iwamoto, K.~Harada, S.~Yamaji and S.~Yoshida, Z. Phys. A {\bf288}, 383 (1978).
\bibitem{arit04}Y.~Aritomo and M.~Ohta, Nucl. Phys. A {\bf744}, 4 (2004)
\bibitem{arit09}Y.~Aritomo, Phys. Rev. C {\bf80}, 064604 (2009)
\bibitem{bloc78}J.~Blocki, et al., Ann. Phys {\bf113}, 330 (1978).
\bibitem{nix84}J.R.~Nix and A.J.~Sierk, Nucl. Phys. A {\bf428}, 161c (1984).
\bibitem{rand84}J.~Randrup and W.J.~Swiatecki, Nucl. Phys. A {\bf429}, 105 (1984).
\bibitem{feld87}H.~Feldmeier, Rep. Prog. Phys {\bf50}, 915 (1987).
\bibitem{carj86}N.~Carjan, A.J.~Sierk, and J.R.~Nix, Nucl. Phys. A {\bf452}, 381 (1986).
\bibitem{carj92}M.~Carjan, et al., AIP Conference Proceeding No.250 (AIP, New York, 1992).
\bibitem{wada93}T.~Wada, Y.~Abe and N.~Carjan, Phys. Rev. Lett {\bf70}, 3538 (1993).
\bibitem{asan06}T.~Asano, T.~Wada, M.~Ohta, S.~Yamaji and H.~Nakahara, J. Nucl. Radio Sci {\bf7}, 7 (2006).
\bibitem{davi76}K.T.R~Davies, A.J.~Sierk and J.R.~Nix, Phys. Rev. C {\bf13}, 2385 (1976).
\bibitem{igna75}A.N.~Ignatyuk, G.N.~Smirenkin and A.S.~Tishin, Sov. J. Nucl. Phys {\bf21}, 255 (1975).
\bibitem{krap79}H.J.~Krappe, J.R.~Nix and A.J.~Sierk, Phys. Rev. C {\bf20}, 992 (1979).
\end{thebibliography}
\end{document}